\def\be{\begin{equation}}
\def\ee{\end{equation}}
\def\bea{\begin{eqnarray}}
\def\eea{\end{eqnarray}}

\documentclass[aps,prl,twocolumn,groupedaddress,amsmath,amssymb,showpacs]{revtex4}
\usepackage[dvips]{graphicx}
\begin{document}

\title{On the Lieb-Liniger model in the infinite coupling constant limit}

\author{St\'ephane Ouvry$^{1,2}$, Alexios P. Polychronakos$^3$}
\email{ouvry@lptms.u-psud.fr,alexios@sci.ccny.cuny.edu}

\affiliation {\makebox{$^1$Universit\'e Paris-Sud, Laboratoire de Physique Th\'eorique et Mod\`eles
Statistiques, B\^at 100, 91405 Orsay Cedex
\\}}

\affiliation {\makebox{$^2$CNRS, LPTMS, Unit\a'e de Mixte de Recherche 8626, 91405 Orsay Cedex, France\\}}

\affiliation {\makebox{$^3$Department of Physics, City College of the CUNY, New York, NY, US\\}}

\date{\today}

\begin{abstract}
We consider the one-dimensional Lieb-Liniger model (bosons interacting via 2-body delta potentials)
in the infinite coupling constant limit (the so-called Tonks-Girardeau model).
This model might be relevant as a description of atomic Bose gases confined in a one-dimensional geometry.
It is known to have a fermionic spectrum since the $N$-body wavefunctions have to vanish at coinciding points,
and therefore be symmetrizations of fermionic Slater wavefunctions. We argue that in the infinite coupling
constant limit the model is indistinguishable from free fermions, i.e., all physically accessible observables
are the same as those of free fermions. Therefore, Bose-Einstein condensate experiments at
finite energy that preserve the one-dimensional geometry cannot test any bosonic characteristic of such a model.
\end{abstract}

\maketitle

{\bf Statistics in one dimension and the hard-core boson model:}
The Lieb-Liniger model of interacting particles on the line is defined as \cite{Lieb}
\be\label{Lieb}
H_N= -{1\over 2}\sum_{i=1}^N({\partial\over\partial x_i})^2+c\sum_{i<j}\delta(x_i-x_j)
\ee
Contact $\delta$-interactions have no effect on fermionic wavefunctions whereas they act nontrivially on bosonic ones.
Such interactions are commonly used to model interacting atomic bosonic gases \cite{Dalibard} and can also be relevant in describing other
aspects of atomic quantum fluids such as, for example, supersolidity \cite{Thouless}.

The Lieb-Liniger model is solvable by Bethe Ansatz techniques for any finite coupling $c$. In the $c\to\infty$
hard core limit, the spectrum of the model becomes fermionic since the eigenstates have to vanish at coincidence points.
On the other hand the statistics of the model, conventionally defined through the symmetry of the wavefunction,
remains bosonic. The energy eigenfunctions $\psi_{n,G}$ are symmetrizations of the fermionic ones $\psi_{n,F}$,
\be
\psi_{n,G} (x_1,x_2,...,x_N) = \prod_{i<j} sgn(x_i - x_j ) \psi_{n,F} (x_1,x_2,...,x_N)
\label{psi}
\ee
and the energy eigenvalues are identical to the fermionic ones. This particular Bose-Fermi mapping has been first noticed by Girardeau \cite{Girardeau}.
The question we address in this note is the exact fermionic nature of the Lieb-Liniger model in this singular limit.

In one dimension, the configuration space for $N$ particles decomposes into $N!$ distinct sectors, according to the
ordering of the particle coordinates. If the particles are identical and indistinguishable, these sectors are all
physically equivalent. In principle we could keep only one sector, the other ones representing ``gauge" (unphysical)
copies.

Consider an $N$-body quantum Hamiltonian and an initial wavefunction with support inside a particular sector
of the configuration space, say, $x_1<x_2<...<x_N$.
For a Hamiltonian without singular interactions, the quantum
time evolution of the wavefunction is such that different sectors end up communicating with each other; that is,
the wavefunction will eventually spread over the remaining sectors. The different sectors interfere and it is
necessary to determine the value of the wavefunction in all of them to fully fix the dynamics.

However, when the interaction becomes
sufficiently singular at coincidence points, tunneling between different sectors might become forbidden, implying that
the wavefunction time evolution is restricted to the initial sector. In such a case, wavefunctions from different
sectors never interfere and live independent lives. The sectors have become effectively superselected and particle
statistics have become irrelevant: we may restrict the wavefunction in one sector, or continue it in a symmetric
or antisymmetric way in the other sectors without affecting the physics.

This situation is precisely encountered in the Calogero model where the singular nature of the
${g(g-1)/(x_i-x_j)^2}$ Calogero interaction prohibits quantum tunneling between different sectors which therefore do not
communicate. The inverse square potential is quantum mechanically impenetrable in such a way that if the particles
are in a given sector among the $N!$ possible, they will stay in this sector for ever. This property of the
Calogero interactions is an important component in understanding the excusion statistics pertaining to the model \cite{Polychronakos}.

In the hard core Lieb-Liniger model the situation is analogous: because of the singular nature of the $\delta$
interactions in the infinite coupling constant limit particle penetration is suppressed. Indeed, the tunneling amplitude for
a particle of momentum $p$ scattering off a delta-potential of
strength $c$ is
\be
T = \frac{ip}{ip-c}
\ee
and tunneling for momenta lower than $c$ is suppressed. In the limit
$c \to \infty$ all tunneling is suppressed.

In the reference sector
$x_1<x_2<...<x_N$ the hard core Lieb-Liniger wave functions $\psi_G$ coincide with fermionic Slater wavefunctions
$\psi_F$. Since the original statistics of the model is assumed to be bosonic, one may continue the wavefunctions
to other sectors of the configuration space in a symmetric way as in (\ref{psi}).
But this continuation is physically irrelevant since, as already
stated, tunneling to the other sectors is impossible. What matters for physics is the wavefunction in the reference
sector. As this is the same as that of free fermions, the system is entirely fermionic.
It follows that the model, which decribes free bosons when $c=0$,
ends up describing free fermions in the $c\to\infty$ limit.

For exchange or fractional statistics models \cite{Ouvry}, such as the two-dimensional anyon model or the one-dimensional
Calogero model, statistical interactions are parametrized by a dimensionless statistical
coupling constant --in the anyon model it is the fractional part of the flux carried by the particles in unit of the flux quantum; in the Calogero model it is $g$. This absence of a scale is natural, as one does not expect
statistical considerations to arise from a dimensionful interaction, which would introduce an additional physical scale.
In the Lieb-Liniger model the coupling constant $c$ has dimensions of inverse length, so it is not a statistical model
per se except when trivially $c=0$, i.e. pure bosons, and when less trivially $c=\infty$, i.e. pure fermions. At
both points the dimensionful scale $c$ indeed disappears from the model. In between one can still view the Lieb-Liniger model as some sort of statistical model
with an energy-dependent statistical parameter, which for energies much higher than $c^2$ is essentially
bosonic, while for energies much lower than $c^2$ is essentially fermionic.

An alternative point of view for the statistics of particles
in one spatial dimension is to restrict the wavefunction from the outset in one sector (a kind of `gauge fixing').
Since each sector has a boundary (where any two particle coordinates coincide) the issue of boundary conditions
on the wavefunction becomes relevant \cite{NRbos}. Neumann ($\partial_n \psi_b =0$) or Dirichlet ($\psi_b =0$) are obvious choices
that preserve hermiticity and are equivalent to symmetric (bosonic) or antisymmetric (fermionic) continuations
of the wavefunction over the remaining sectors, respectively. Linear combinations of the above of the sort
$\partial_n \psi_b + c \psi_b =0$ are also possible, and correspond to the bosonic Lieb-Liniger model of strength
$c$. Once again, we recover the case $c=\infty$ as corresponding to fermionic statistics.

{\bf Physical observables and momentum:}
One crucial aspect that makes the Tonks-Girardeau gas apparently distinct from free fermions is its momentum distribution.
As the wavefunctions of the two systems differ only by their phase (sign) in each sector, coordinate observables
have the same expectation value,
but the momentum wavefunctions (Fourier transforms) are genuinely different \cite{Lenard}. How can this be reconciled with the
statement that the two models are physically equivalent?

To answer this we must address the question of what are physically accessible observables in the Tonks-Girardeau system.
Concentrating on the momentum, any local interaction of the system with a measuring device would involve a
fully symmetric polynomial of finite degree in the particle momenta $p_1 , \dots , p_N$.

To reduce the problem to its bare bones, we examine the simplest stituation of two particles. The center of
mass momentum $p_1 + p_2$ is an observable, as well as all its powers. The relative momentum
$p = (p_1 - p_2 )/2$, on the other hand, is not an observable since it is not invariant under particle permutations,
but its even powers $p^{2n}$ are.

It is now easy to establish that the center of
mass momentum has the same matrix elements in both models. Indeed, since the particle
interaction in the Tonks-Girardeau model is translation invariant, the center of mass motion is free. We may factor out the
center of mass motion and consider only the relative part, with wavefunction $\psi (x)$, $x = x_1 - x_2$.
For the relative momentum observables $p^{2n}$ we have
\be
< p^{2n} > = \int dx \psi^* (-i \partial_x)^{2n} \psi = \int dx |\partial_x^n \psi |^2
\label{expp}
\ee
where we integrated by parts $n$ times and assumed the usual regular behavior of $\psi$ at $x \to \pm \infty$.
Since $\psi_G (x) = sgn(x) \psi_F (x)$ and both wavefunctions vanish
linearly at $x=0$ we have
\be
\partial_x \psi_G = sgn(x) \partial_x \psi_F ~,~~~
\partial_x^n \psi_G \sim \delta^{(n-1)} (x) ~,~ n\ge 2
\label{part}
\ee
That is, $\psi_G$ has a discontinuity in the first derivative and thus
develops as the leading singular term a delta function of order $n-1$ at $x=0$ in the $n$-th
derivative.

>From (\ref{expp}) and (\ref{part}) we conclude that
\be
< p^2 >_G = < p^2 >_F ~,~~~ < p^{2n} >_G = \infty ~,~ n \ge 2
\ee
Therefore, all observables that would be different between the two models have infinite expectation value in
the Tonks-Girardeau model and thus are not physically measurable, requiring an infinite energy to do so. The physical reason
behind this effect is that such operators, added as perturbations to the system's Hamiltonian due to the
coupling to the measuring device, have the effect of ``pushing'' particles through each other and the hard-core
potential between particles imposes an infinite energy penalty to this process.

In a similar way we can show that mixed local operators of the form $x^m p^{2n-m}$ (or their hermitian symmetrization)
follow a similar pattern:
\bea
< x^m p^{2n-m} >_G &=& < x^m p^{2n-m} >_F ~,~~~ n-m \le 1 \cr
< x^m p^{2n-m} >_G &=& \infty ~,~~~ n-m \ge 2
\eea
That is, operators that are not the same as in the fermion system have infinite expectation value and are, thus,
inaccessible.

There are, of course, other $p$-dependent operators that are different between the two models and do not
have infinite expectation value, such as, e.g., $\cos (ap)$ with $a$ a constant. These operators,
however, are nonlocal and have the property to ``teleport'' particles over each other, thus avoiding the
interaction region and the corresponding energy cost. (The two-particle permutation operator itself is
another example.) Such observables are not physically realizable. We can argue that operators mixing
the superselection sectors manipulate gauge copies of the configuration space and are, in general,
unphysical.

{\bf Time evolution and momentum operator:}
From the discussion of the preceding section it becomes clear that operators such as $p^4$ become unphysical
for the Tonks-Girardeau model and therefore the momentum operator itself is unsuitable. The problem
is similar to the quantum mechanical description of a particle on the half-line. It is known that the standard
operators $x$ and $p$ are unsuitable, and a minimal
set of physical operators is $x^2$, $p^2$ and $xp+px$, closing into the $SL(2)$ algebra. The Casimir of this
algebra fixes the physical Hilbert space and implies that $p$ may not be represented as $-i\partial_x$ on
this space.

To identify an appropriate momentum operator we prefer here to follow
a more physical approach and go back to basics, remembering the justification for
defining $-i\partial_x$ in the standard Schrodinger picture as the momentum operator or, equivalently, the
Fourier transform of the coordinate wavefunction as the momentum wavefunction. We shall work again with
the simplest case of two particles and factor out the center of mass coordinate, leaving us with the
relative coordinate $x$.

Consider a free system with initial wavefunction in the vicinity of $x=0$ and an overall
uncertainty in the position $x$ of order $\Delta x$. We let the system evolve for a long time $T$
and measure the position $x$.
Typically $x$ will have some large value $L$, as the particles have moved away from each other.
If we measure $x$ to be in the range $[L-\Delta L,L+\Delta L]$ we know that it moved a distance $D = x_f - x_i =
L \pm \Delta L \pm \Delta x$ in time $T$.
Its speed, and correspondingly the momentum (assuming unit reduced mass), is estimated then as
\be
p = v = \frac{L}{T} \pm \frac{\Delta L}{T} \pm \frac{\Delta x}{T}
\ee
By taking $T$ large enough, and scaling $L$ and $\Delta L$ with $T$,
we can make the uncertainty of $p$ due to the initial $\Delta x$ negligible.
The probability of the momentum, then, being within a range $\Delta p$ around the value $p$ is the probability
of measuring the coordinate $x$ within a range $\Delta L=T \Delta p$ around the value $L=Tp$, for $T \to \infty$.

The large-$T$ evolution of the wavefunction can be calculated from the initial wavefunction and the free
evolution operator
\be
U(T) = e^{-iHT} = e^{i T\partial_x^2 /2}
\ee
In terms of the Fourier transform of the wavefunction $\phi(k)$ we have
\be
\psi(x,T) = \frac{1}{\sqrt{2\pi}} \int dk e^{-iT k^2 /2 + i k x} \phi(k)
\ee
Putting $x=Tp$ and performing a standard saddle-point expansion for large $T$ we obtain
\be
\psi(Tp,T) = \frac{1}{\sqrt T} \phi(p) ~~(T \to \infty)
\ee
Since the probability for measuring $x$ in a range $T \Delta p$ around $Tp$ is $T \Delta p | \psi (Tp,T) |^2$,
we find the probability for measuring the momentum in the range $\Delta p$ around $p$ to be
\be
P(p,p+\Delta p) = \Delta p |\phi(p)|^2
\ee
which justifies the role of the Fourier transform $\phi(p)$ as the momentum wavefunction.

We can now use this procedure to define the momentum of particles in the Tonks-Girardeau system: let the particles
fly away from each other. The distribution of their positions after they spread enough will be taken to represent their kinematical
velocity, or momentum, distribution. Although this
evolution is not really free, interactions occur only at
coincidence points and are an integral part of the definition of the
system, just like the restriction to $x>0$ for the particle on the
half-line.

It should be clear that this distribution will be {\it identical to the one of
freely moving fermions}. To see this, use the fact that the spectra of the two models are identical and that
the energy eigenfunctions are related as in (\ref{psi}). Then, expanding the initial wavefunction in
terms of energy eigenstates, we have
\bea
\psi_{_G} (x_1,...,x_N) &=& \sum_n c_n \psi_{n,G} (x_1,...,x_N) \cr
&=& \sum_n c_n \prod_{i<j} sgn(x_i - x_j ) \psi_{n,F} (x_1,...,x_N) \cr
&=& \prod_{i<j} sgn(x_i - x_j ) \sum_n c_n \psi_{n,F} (x_1,...,x_N) \cr
&=& \prod_{i<j} sgn(x_i - x_j ) \psi_F (x_1,...,x_N)
\eea
In other words, the hard-core wavefunction and the corresponding fermionic wavefunction have the same
energy eigenvalue expansion coefficients. Thus, at any time $t$,
\bea
\psi_{_G} (x_1,...,x_N, t) &=& \sum_n c_n e^{-i E_n t} \psi_{n,G} (x_1,...,x_N) \cr
&=& \prod_{i<j} sgn(x_i - x_j ) \psi_F (x_1,...,x_N,t)\nonumber\\
\eea
So the wavefunctions of the hard-core boson model and the free fermion model remain at all times related
by a simple sign change over sectors. Therefore the particle distribution after a large time $T$, and so the
induced momentum distribution, are that of free fermions.

For general interacting systems, one can recover the standard momentum distribution by
{\it switching off the interactions}
and then letting the particles fly away. This would reproduce the (bosonic) Fourier transform as the momentum distribution.
The arguments of the preceding
sections assume the permanent presence of the infinite-strength interactions. Our point is that any real fly-away experiment
{\it preserving the one-dimensional geometry and interactions}
would gather
data consistent with an initial free fermion momentum distribution.

The above can also be summarized by saying that a complete set of physical observables is the set of
all {\it time-dependent} amplitudes of the form
\be
\left| <x_1 , \dots , x_N ; t | x'_1 , \dots , x'_N ; t' > \right|^2 =
\left| G(x'_i - x_i ; t'-t ) \right|^2
\ee
From the present analysis it is clear that these amplitudes for the Tonks-Girardeau model
are identical to those of free fermions.

{\bf Conclusions:}
We have argued that the Tonks-Girardeau model is physically
indistinguishable from free fermions, on the basis that any
experiment aiming to distinguish the two models while staying within the one-dimensional setting
would require an infinite energy or the action of unphysical operators.
Inferring particle momenta from the
fly-away spatial distribution of Tonks-Girardeau particles
would return the distribution of free fermions.

The above considerations are, in fact, quite relevant to the experimental
setup decribed in \cite{Kinoshita,Kinobis}:
allow the Tonks-Girardeau atoms, originally constrained in a trap,
to fly away ballistically while they are still in a 1d geometry and use the asymptotic particle
distribution to deduce the momentum distribution. As argued above,
such an experiment would fail to detect any bosonic features of the system.

In experimental realizations of the model the interactions are not
entirely hard-core; that is, the strength of the potential will not
be infinite. For a finite $c$ the tunneling between sectors is not
suppressed for momenta $p>c$ and the energy cost of measurements
that would distinghuish the model from fermions is not infinite.
In effect, the model describes free fermions for momenta $p\ll c$ and
free bosons for momenta $p \gg c$, with a crossover region in between.
Interpretation of experimental data, therefore, must take into
account the finite strength of the core repulsion relative to the
momentum scale.

The arguments presented in the previous analysis can be repeated
and verified in the discrete
lattice version of the model, that is, the Bose-Hubbard
model \cite{Hubbard} with infinite on-site repulsion energy and
finite hopping parameter. In this model at most one boson is allowed
on a given site, so again the particles are fermion-like.
The bosonic and fermionic Hilbert spaces are in exact one-to-one
correspondence via a Jordan-Wigner transformation
that maps each bosonic
operator into a fermionic one, and vice versa. The total
momentum operator is common in the two descriptions, but
the relative momentum operators differ: the bosonic one
may create states of infinite energy (since it may move
two bosons on top of each other), while the fermionic one does not.
Again, we recover the same basic conclusions as in the
continuous model.

{\it \underline{Acknowledgements:}} S.O. thanks Gora Shlyapnikov for discussions and Nigel Cooper for interesting comments on the paper and for drawing our
attention to \cite{Kinobis}. A. P. thanks Universit\'e Paris-Sud for supporting his stay at LPTMS as
Professeur Invit\'e. We also thank the Galileo Galilei Institute for Theoretical Physics for the hospitality
and the INFN for partial support during the completion of this work.

\end{document}